\pdfoutput=1
\documentclass[%
reprint, 
superscriptaddress, 
 amsmath,amssymb,
 aps,
 pra,
]{revtex4-2}

\usepackage{graphicx}
\usepackage{dcolumn}
\usepackage{bm}

\begin{document}

\title{Chiral cavity-magnonic system for the unidirectional photon blockade}

\author{Jiaxin Yang}
\email{yangjx@snnu.edu.edu}
\affiliation{School of Physics and Information Technology, Shaanxi Normal University, Xi’an 710119, China}
\affiliation{Xi’an Key Laboratory of Optical Information Manipulation and Augmentation, Xi’an 710119, China}

\author{Yilou Liu}
\affiliation{School of Physics and Information Technology, Shaanxi Normal University, Xi’an 710119, China}
\affiliation{Xi’an Key Laboratory of Optical Information Manipulation and Augmentation, Xi’an 710119, China}

\author{Rui-Shan Zhao}
\affiliation{School of Physics and Information Technology, Shaanxi Normal University, Xi’an 710119, China}

\author{Xiao-Tao Xie}
\email{xtxie@snnu.edu.edu}
\affiliation{School of Physics and Information Technology, Shaanxi Normal University, Xi’an 710119, China}
\affiliation{Xi’an Key Laboratory of Optical Information Manipulation and Augmentation, Xi’an 710119, China}

\date{\today}

\begin{abstract}
We propose an scheme for directional single-photon source based on a chiral cavity-magnon system. In this system, the magnon mode in a single-crystal yttrium iron garnet (YIG) sphere is coupled to only one of two rotating microwave modes in the torus-shaped microwave cavity. When two-photon drives are applied to both ports of the waveguide, the chiral cavity–magnon coupling leads to an unconventional photon blockade in one propagation direction, resulting in directional single-photon emission. The emission direction of the single-photon source can be controlled by reversing the biased magnetic field. Furthermore, we further examine the effects of imperfect chiral cavity–magnon coupling and the coupling between the two cavity modes on the photon blockade behavior. The results show that the system retains robustness in the presence of these nonideal factors, and the unidirectional photon blockade effect remains clearly preserved. Our approach may offer a new perspective for the design of unidirectional single-photon devices.
\end{abstract}

                                                            
\maketitle

\section{Introduction}
Chirality, a fundamental manifestation of asymmetry in nature, plays a crucial role across chemistry, optics, and quantum physics \cite{penasa2025advances, haupert2009chirality, yan2024structural, wang2024chiral}. In the field of quantum optics, the introduction of chirality breaks time-reversal symmetry, enabling nonreciprocal transmission and directional energy flow \cite{jalas2013and, mitsch2014quantum, coles2016chirality}. This unique physical effect has made chiral quantum optics an emerging research focus in areas such as quantum communication, quantum computing, and on-chip quantum devices \cite{lodahl2017chiral, petersen2014chiral, sollner2015deterministic, PhysRevLett.117.240501, tang2022chiral}. In nanophotonic structures, such as waveguides and whispering-gallery-mode (WGM) resonators \cite{coles2016chirality, PhysRevLett.110.213604, PhysRevLett.126.233602}, chiral light–matter interactions are particularly pronounced, providing an ideal platform for directional control of photons. Leveraging this chiral mechanism, researchers have proposed a variety of quantum functionalities, including photon entanglement \cite{PhysRevA.94.012302, PhysRevB.92.155304}, photon squeezing \cite{PhysRevA.108.033701}, and quantum droplet \cite{8nvh-541r}, among which directional emission has attracted significant attention due to its important applications in quantum information transfer and the development of quantum light sources \cite{PhysRevA.99.043833, PhysRevA.105.063719, PhysRevA.106.043722, PhysRevB.95.121401, sun2019separation}. At present, a large number of chiral devices have been proposed both theoretically and experimentally, such as isolators \cite{PhysRevLett.121.203602, PhysRevX.5.041036, PhysRevA.90.043802}, circulators \cite{scheucher2016quantum, PhysRevLett.121.203602}, single-photon routing \cite{shomroni2014all, PhysRevA.97.023821, PhysRevA.94.063817}, and single-photon diodes \cite{PhysRevA.98.043852, scheucher2016quantum}. 

Photon blockade (PB) is one of the fundamental quantum mechanisms for realizing single-photon sources, and its concept was originally proposed by analogy with the Coulomb blockade \cite{PhysRevLett.79.1467}. According to the underlying physical mechanisms, photon blockade can be broadly categorized into two types \cite{PhysRevLett.134.183601}: conventional photon blockade (CPB) \cite{PhysRevLett.109.193602, PhysRevA.87.023809} and unconventional photon blockade (UPB) \cite{PhysRevLett.104.183601, PhysRevA.83.021802}. The former relies on a strong nonlinearity that modifies the system’s energy-level structure, such that when the driving field is tuned to resonance with the transition between the vacuum and single-photon states, the excitation of multiple photons is effectively suppressed \cite{PhysRevA.89.043818, PhysRevLett.125.197402, PhysRevA.109.043702}. CPB has been experimentally demonstrated in various physical platforms, including circuit quantum electrodynamics (QED) systems \cite{PhysRevLett.107.053602, PhysRevLett.106.243601}, quantum dots embedded in photonic crystals \cite{faraon2008coherent, PhysRevLett.114.233601}, and optical cavities containing trapped atoms \cite{birnbaum2005photon, PhysRevLett.118.133604}. In contrast, the latter is induced by destructive quantum interference among different driven–dissipative pathways and, unlike CPB, can occur even in the weakly nonlinear regime \cite{PhysRevA.96.053810, PhysRevLett.127.240402, PhysRevA.88.033836, PhysRevLett.129.043601}. UPB has also been experimentally observed in quantum dot–cavity QED systems \cite{PhysRevLett.121.043601} and in two coupled superconducting resonators \cite{PhysRevLett.121.043602}. Furthermore, many quantum systems have been theoretically predicted to exhibit photon blocking effects, such as optomechanical systems \cite{PhysRevLett.107.063601, PhysRevA.88.023853,PhysRevLett.114.093602, PhysRevA.87.013839, Liu:23, Wang:23}, the Tavis-Cummings system \cite{st87-3cxz}, and hybrid antiferromagnet-cavity quantum systems \cite{m2rq-l6fl}. In recent years, nonreciprocal photon blockade \cite{PhysRevLett.121.153601, li2019nonreciprocal} has attracted increasing attention as a novel quantum regulation mechanism.

In a recent study, a yttrium iron garnet (YIG) microsphere was placed at a specific position inside a torus-shaped microwave cavity, and its ferromagnetic resonance was tuned to the transverse-electric (TE) cavity mode, leading to the realization of circularly polarized unidirectional cavity magnon polaritons \cite{PhysRevApplied.19.014030}. At this particular position, the clockwise (CW) and counterclockwise (CCW) cavity modes both exhibit circular polarization, allowing only one mode to couple effectively with the magnet. Building on the above study, we propose a theoretical model to realize directional photon blockade in a chiral cavity–magnon system. When two-photon driving fields are simultaneously applied from both ports of the waveguide, the system exhibits nonreciprocal quantum statistical behavior: two-photon emission occurs only from one side, while the opposite side is restricted to single-photon emission due to the photon blockade effect. This mechanism enables a directional conversion from two photons to one photon, providing a new approach for implementing nonreciprocal single-photon sources and quantum information transfer devices.

This work is organized as follows. In Sec.~\ref{Sec2}, we introduce the physical model of the system and the calculation of the optimal conditions for photon blockade. Section~\ref{Sec3} presents the results of directional photon blockade and analyzes its robustness. A brief summary is given in Sec.~\ref{Sec4}.

\section{Model}\label{Sec2}
\begin{figure}
	\includegraphics[width=8.6cm, height=6.158cm, clip]{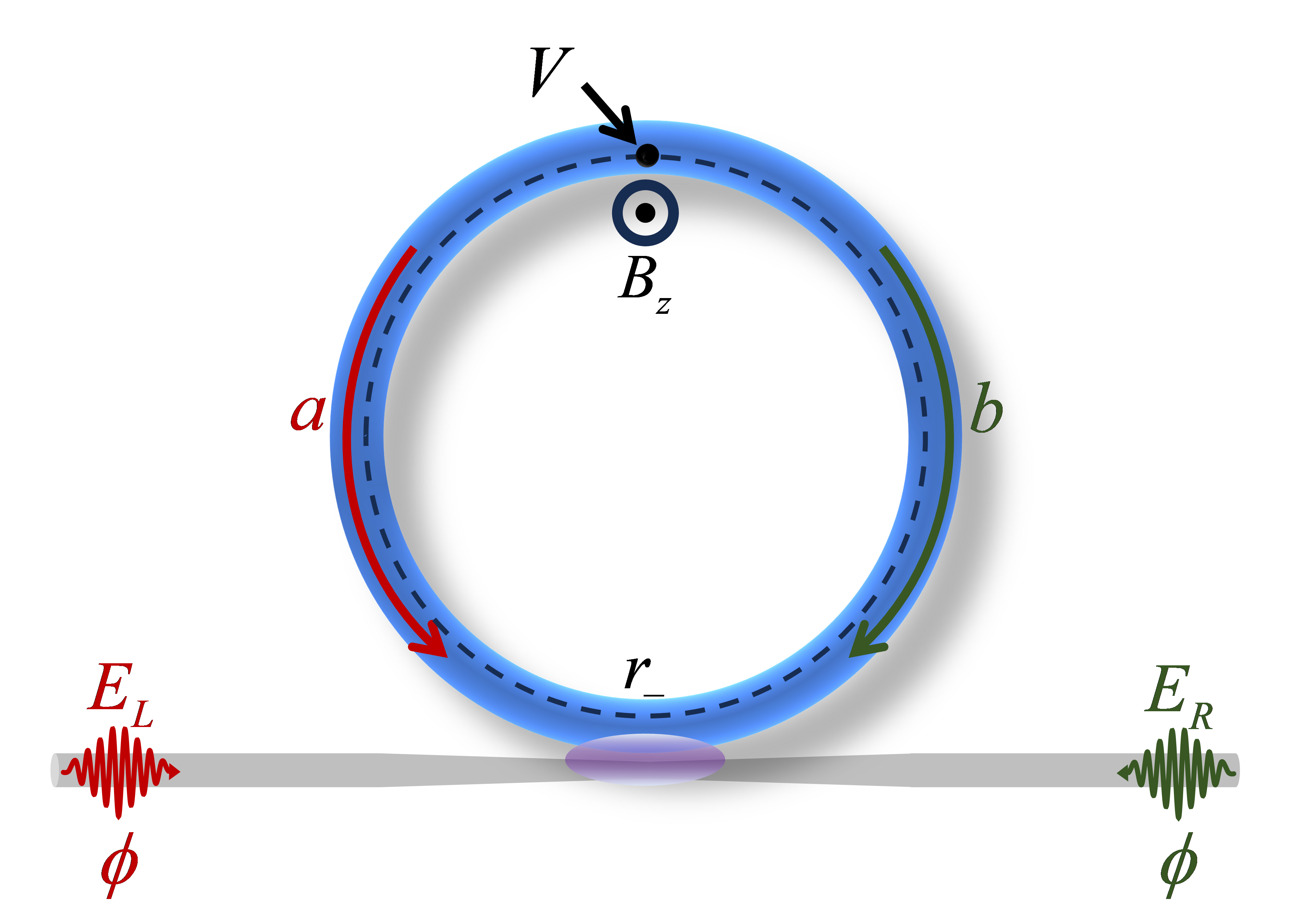}
	\caption{(Color online) Schematic illustration of the hybrid cavity-magnon system. A YIG sphere is placed at a specific position of thetorus-shaped microwave cavity and is biased by a static magnetic field $B_z$ applied perpendicular to the ring plane. The dashed lines indicate the special radial positions $r_{-}$. The sphere is additionally subjected to an external drive $V$. The waveguide is evanescently coupled to the cavity and is driven from both ports by two-photon fields $E_L$ and $E_R$ with phase $\phi$. The cavity supports a pair of degenerate CCW $(a)$ and CW $(b)$ rotating microwave modes.} \label{Model}
\end{figure}

As shown in Fig.~\ref{Model}, we study a torus-shaped microwave cavity with a YIG sphere inside and waveguide coupled hybrid cavity–magnon system. When driving from both ports of the waveguide, CCW mode and CW mode are excited in the cavity. The two photon modes are backscattered by the medium of the sphere or port to produce a weak coupling with a coupling strength of $J$. The total Hamiltonian of the system is written as (we set $\hbar$ = 1 hereafter)
\begin{equation}
\begin{aligned}
	\hat H =& {\omega _c}{\hat a^\dag }\hat a + {\omega _c}{\hat b^\dag }\hat b + {\omega _m}{\hat m^\dag }\hat m + J({\hat a^\dag }\hat b + {\hat b^\dag }\hat a) \\&+ {g_a}({\hat a^\dag }\hat m + {\hat m^\dag }\hat a) + {g_b}({\hat b^\dag }\hat m + {\hat m^\dag }\hat b)  \\&+ {E_L}({e^{i(\phi  - {\omega _e}t)}}{{\hat a}^{\dag 2}} + {e^{ - i(\phi  - {\omega _e}t)}}{{\hat a}^2}) \\&+ {E_R}({e^{i(\phi  - {\omega _e}t)}}{{\hat b}^{\dag 2}} + {e^{ - i(\phi  - {\omega _e}t)}}{{\hat b}^2})\\&+ V({e^{ - i{\omega _o}t}}{{\hat m}^\dag } + {e^{i{\omega _o}t}}\hat m),\label{H_a}
\end{aligned}
\end{equation}
where $\omega _c$ and $\omega _m$ denote the resonance frequencies of the cavity and magnon modes, respectively, $a$, $b$, and $m$ represent annihilation operators for the CCW cavity mode, the CW cavity mode, and the magnon mode, respectively. The chiral coupling strengths between the two cavity modes and the magnons are represented by $g_a$ and $g_b$, respectively. In this system, an external bias magnetic field $B_z$ is applied perpendicular to the plane of the cavity to magnetize the YIG sphere. The magnetized YIG supports either right-handed or left-handed magnon precession, depending on the polarity of the magnetic field. Meanwhile, the cavity supports two degenerate omicrowave modes propagating in the CW and CCW directions, which carry opposite orbital angular momenta. Since the photon–magnon interaction must conserve angular momentum, only the cavity mode with a rotational sense matching that of the magnon can effectively couple to it, while the other mode remains decoupled. Consequently, the system exhibits chiral cavity–magnon coupling \cite{PhysRevApplied.19.014030, PhysRevB.102.064416, PhysRevApplied.13.044039, PhysRevApplied.16.064066, PhysRevA.101.043842}, where the coupling direction can be switched between the CW and CCW cavity modes by reversing the magnetic field direction, i.e. ${g_a} \ne 0$ and ${g_b} = 0$ or ${g_a} = 0$ and ${g_b} \ne 0$. ${E_\alpha }(\alpha=L, R)$, $\omega_e$ and $\phi$ characterize the strength, frequency and phase of the parametric driving in the two driving directions, respectively. A weak probe field $V$ with frequency $\omega_o$ is used to drive the magon mode.

Under the frequency-matching condition ${\omega _e} = 2{\omega _o}$, and applying the rotating frame defined by $\hat U = \exp [ - i{\omega _o}({{\hat a}^\dag }\hat a + {{\hat b}^\dag }\hat b + {{\hat m}^\dag }\hat m)t]$, the transformed total Hamiltonian can be rewritten as
\begin{equation}
	\begin{aligned}
		\hat H_r =& {\Delta _c}{{\hat a}^\dag }\hat a + {\Delta _c}{{\hat b}^\dag }\hat b + {\Delta _m}{{\hat m}^\dag }\hat m + J({{\hat a}^\dag }\hat b + {{\hat b}^\dag }\hat a) \\&+ {g_a}({{\hat a}^\dag }\hat m + {{\hat m}^\dag }\hat a) + {g_b}({{\hat b}^\dag }\hat m + {{\hat m}^\dag }\hat b) \\&+ {E_L}({e^{i\phi }}{{\hat a}^{\dag 2}} + {e^{ - i\phi }}{{\hat a}^2}) \\&+ {E_R}({e^{i\phi }}{{\hat b}^{\dag 2}} + {e^{ - i\phi }}{{\hat b}^2})\\& + V({{\hat m}^\dag } + \hat m),\label{H_b}
	\end{aligned}
\end{equation}
where $\Delta_c=\omega_c-\omega_o$ and $\Delta_m=\omega_m-\omega_o$ correspond to the
detunings of the cavity field and the magnon with respect to the driving laser, respectively.

The effective non-Hermitian Hamiltonian of the system can be obtained by phenomenologically adding imaginary dissipation terms into Hamiltonian  Eq.~(\ref{H_b}), and is given by
\begin{equation}
	\begin{aligned}
	{{\hat H}_{{\rm{eff}}}} = {{\hat H}_r} - i\frac{{{\kappa_a}}}{2}{{\hat a}^\dag }\hat a - i\frac{{{\kappa_b}}}{2}{{\hat b}^\dag }\hat b - i\frac{{{\kappa_m}}}{2}{{\hat m}^\dag }\hat m,\label{H_eff}
	\end{aligned}
\end{equation}
where $\kappa_a(\kappa_b)$ and $\kappa_m$ are the dissipation rates of the CCW(CW) mode and magnon mode, respectively. For the sake of simplicity, we assume $\kappa_a=\kappa_b=\kappa_c$ and ${E_L}={E_R}=E$ in the following discussion.

In the weak-driving-field limit $E$ and $V$ $\ll$ $\kappa_c$, ${\kappa _m}$, the Hilbert space of the system can be approximately truncated within
the two excitation subspaces. Therefore, the wave function $\left| \psi  \right\rangle$ of the system can be expressed as
\begin{small}
	\begin{equation}
	\begin{aligned}
		\left| \psi  \right\rangle  =& {C_{000}}\left| {000} \right\rangle  + {C_{100}}\left| {100} \right\rangle  + {C_{010}}\left| {010} \right\rangle  + {C_{001}}\left| {001} \right\rangle  \\&+ {C_{110}}\left| {110} \right\rangle  + {C_{101}}\left| {101} \right\rangle  + {C_{011}}\left| {011} \right\rangle  \\&+ {C_{200}}\left| {200} \right\rangle  + {C_{020}}\left| {020} \right\rangle  + {C_{002}}\left| {002} \right\rangle ,\label{psi}
	\end{aligned}
\end{equation}
\end{small}
with probability amplitudes ${C_{amb}}$. Substituting Eqs.~(\ref{H_eff}) and (\ref{psi}) into the non-Hermitian Schr\"{o}dinger equation $i\partial \left| \psi  \right\rangle /\partial t = {{\hat H}_{{\rm{eff}}}}\left| \psi  \right\rangle $, we can obtain a set of differential equations for the probability amplitudes:
\begin{small}
\begin{equation}
\begin{aligned}		
		i{{\dot C}_{000}} &= V{C_{010}} + \sqrt 2 E{e^{ - i\phi }}{C_{200}} + \sqrt 2 E{e^{ - i\phi }}{C_{002}},\\
		i{{\dot C}_{100}} &= {{\tilde \Delta }_c}{C_{100}} + J{C_{001}} + {g_a}{C_{010}} + V{C_{110}},\\
		i{{\dot C}_{010}} &= {{\tilde \Delta }_m}{C_{010}} + {g_a}{C_{100}} + {g_b}{C_{001}} + V{C_{000}}+ \sqrt 2 V{C_{020}},\\
		i{{\dot C}_{001}} &= {{\tilde \Delta }_c}{C_{001}} + J{C_{100}} + {g_b}{C_{010}} + V{C_{011}},\\
		i{{\dot C}_{110}} &= ({{\tilde \Delta }_c} + {{\tilde \Delta }_m}){C_{110}} + J{C_{011}} + \sqrt 2 {g_a}{C_{200}} \\&+ \sqrt 2 {g_a}{C_{020}} + {g_b}{C_{101}} + V{C_{100}},\\
		i{{\dot C}_{101}} &= 2{{\tilde \Delta }_c}{C_{101}} + \sqrt 2 J{C_{200}} + \sqrt 2 J{C_{002}} + {g_a}{C_{011}} \\&+ {g_b}{C_{110}},\\
		i{{\dot C}_{011}} &= ({{\tilde \Delta }_c} + {{\tilde \Delta }_m}){C_{110}} + J{C_{110}} + \sqrt 2 {g_b}{C_{002}} \\& + \sqrt 2 {g_b}{C_{020}} + {g_a}{C_{101}} + V{C_{001}},\\
		i{{\dot C}_{200}} &= 2{{\tilde \Delta }_c}{C_{200}} + \sqrt 2 J{C_{101}} + \sqrt 2 {g_a}{C_{110}} + \sqrt 2 E{e^{i\phi }}{C_{000}},\\
		i{{\dot C}_{020}} &= 2{{\tilde \Delta }_m}{C_{020}} + \sqrt 2 {g_a}{C_{110}} + \sqrt 2 {g_b}{C_{011}} + \sqrt 2 V{C_{010}},\\
		i{{\dot C}_{002}} &= 2{{\tilde \Delta }_c}{C_{002}} + \sqrt 2 J{C_{101}} + \sqrt 2 {g_b}{C_{011}} + \sqrt 2 E{e^{i\phi }}{C_{000}},\label{proamp}
\end{aligned}
\end{equation}
\end{small}
where ${{\tilde \Delta }_c} = {\Delta _c} - i\kappa_c /2$ and ${{\tilde \Delta }_m} = {\Delta _m} - i{\kappa _m}/2$. The probability amplitudes can be solved by neglecting the higher-order term in each equation and setting ${{\dot C}_{amb}}=0$ in the steady state. Thus the solutions for the probability amplitudes ${C_{100}}$ and ${C_{200}}$ of the steady state are obtained as follows (more details can be found in Appendix \ref{AppendixA}):
\begin{figure}
	\includegraphics[width=8.6cm, height=6.675cm, clip]{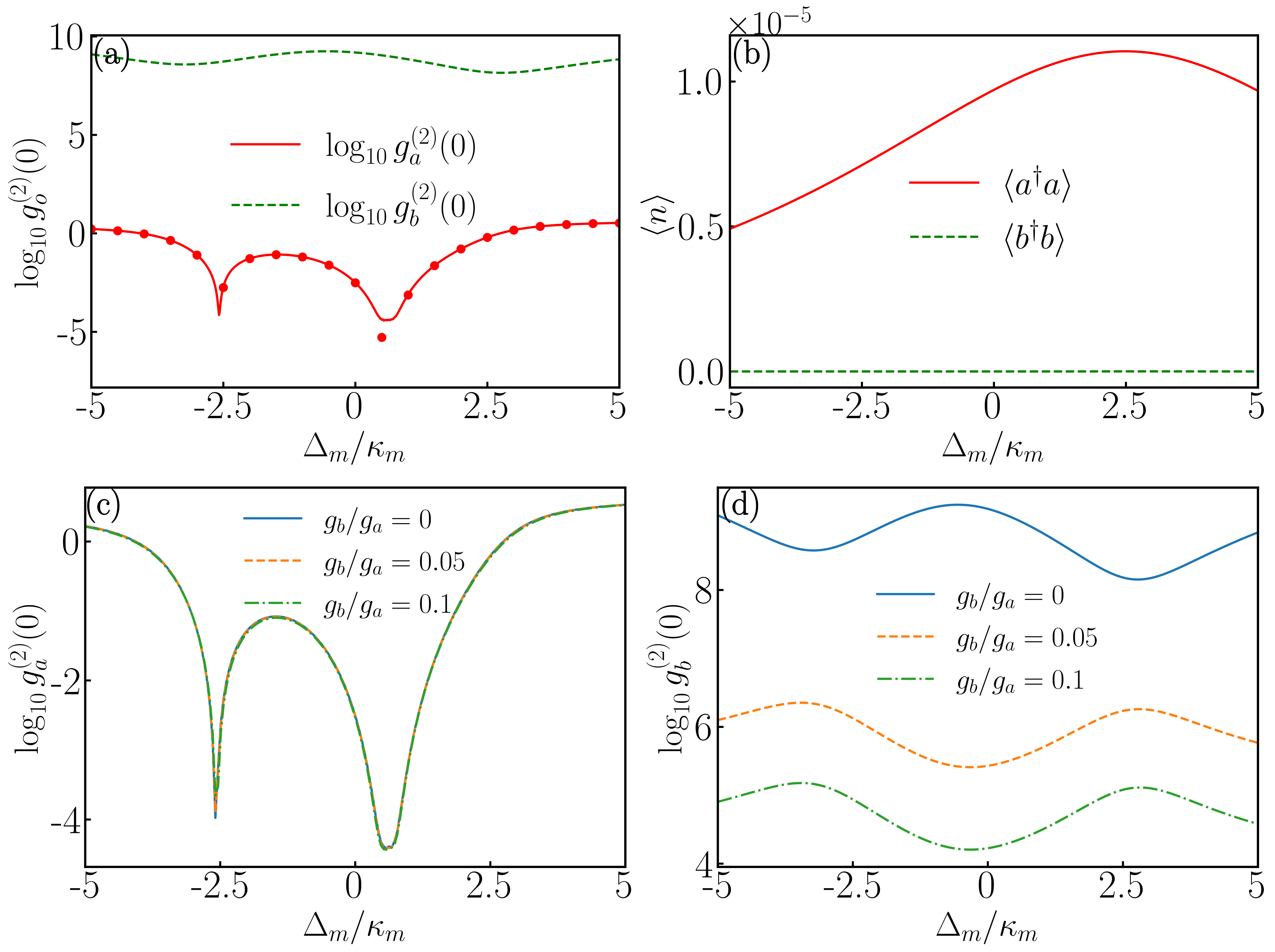}
	\caption{(Color online) (a) Logarithmic plot of the second-order correlation function ${\log _{10}}g_a^{(2)}(0)$ (red solid line) and ${\log _{10}}g_b^{(2)}(0)$ (green dashed line) are plotted as functions of the detuning $\Delta_m/\kappa_m$. The red circle and green triangle indicate the analytical results. (b) The mean photon number of $a$ mode and $b$ mode
as functions of the detuning $\Delta_m/\kappa_m$. Logarithmic plot of the second-order correlation function (c) ${\log _{10}}g_a^{(2)}(0)$ and (d) ${\log _{10}}g_b^{(2)}(0)$ versus the detuning $\Delta_m/\kappa_m$ for the different coupling ratios $g_b/g_a$ = 0, 0.05, and 0.1. Other parameters are $E=E_{{\rm{opt}}}$, $\phi= -0.41\pi$, $\kappa_m=4\pi$MHz, $\kappa_c=2.5\kappa_m$, $J=0$, $g_a=3\kappa_m$, $g_b=0$, $V=0.01\kappa_m$ and $\Delta_c=0.5\kappa_m$.} \label{Fig2}
\end{figure}
\begin{figure*}
	\includegraphics[width=17.2cm, height=4.28cm, clip]{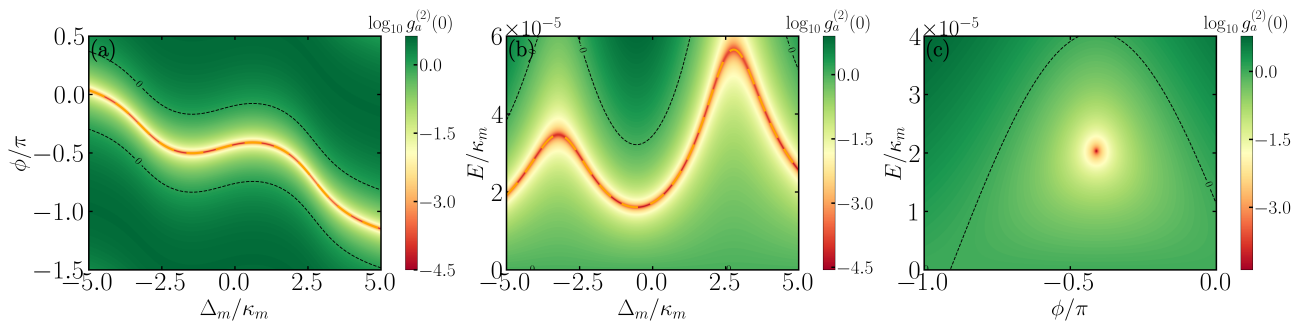}
	\caption{(Color online) Second-order correlation function on a logarithmic scale as a function of (a) the phase $\phi/\pi$, (b) the driving strength $E/\kappa_m$, and the detuning $\Delta_m/\kappa_m$, and (c) both $\phi$ and $E$. The orange dashed lines in (a) and (b) are determined by Eq.~\ref{opt}. In (c)$\Delta_c=\Delta_m=0.5\kappa_m$; all other parameters are the same as in Fig.~\ref{Fig2}.} \label{Fig3}
\end{figure*}
\begin{small}
\begin{equation}
\begin{aligned}
	{C_{100}} &= \frac{{VZ}}{P},\\
	{C_{200}} &= \frac{{{V^2}{Z^2}Q - {E{e^{i\phi }}P({{\tilde \Delta }_c}g_a^4 - 2J{g_b}g_a^3 + {K_2}g_a^2 + {K_1}{g_a} + {K_0})}}}{{\sqrt 2{P^2}}Q},\label{solofproapmp}
\end{aligned}
\end{equation}	
\end{small}
where $Z = {g_b}J - {g_a}{{\tilde \Delta }_c}$; $P =  - 2{g_a}{g_b}J + g_a^2{{\tilde \Delta }_c} + g_b^2{{\tilde \Delta }_c} + ( {J - {{\tilde \Delta }_c}} )( {J + {{\tilde \Delta }_c}} ){{\tilde \Delta }_m}$; $Q = 2{g_a}{g_b}J + g_a^2( {{{\tilde \Delta }_c} + {{\tilde \Delta }_m}} ) + g_b^2( {{{\tilde \Delta }_c} + {{\tilde \Delta }_m}} ) - 2{{\tilde \Delta }_c}( { - J + {{\tilde \Delta }_c} + {{\tilde \Delta }_m}} )( {J + {{\tilde \Delta }_c} + {{\tilde \Delta }_m}} )$; ${K_2} = {J^2}{{\tilde \Delta }_m} + (g_b^2 - {{\tilde \Delta }_c}({{\tilde \Delta }_c} + {{\tilde \Delta }_m}))(2{{\tilde \Delta }_c} + {{\tilde \Delta }_m})$, ${K_1} = 2{g_b}J(g_b^2 + 2\tilde \Delta _c^2)$, ${K_0} = g_b^4({{\tilde \Delta }_c} + {{\tilde \Delta }_m}) - g_b^2({J^2}{{\tilde \Delta }_m} + {{\tilde \Delta }_c}({{\tilde \Delta }_c} + {{\tilde \Delta }_m})(2{{\tilde \Delta }_c} + 3{{\tilde \Delta }_m})) + 2\tilde \Delta _c^2{{\tilde \Delta }_m}( - J + {{\tilde \Delta }_c} + {{\tilde \Delta }_m})(J + {{\tilde \Delta }_c} + {{\tilde \Delta }_m})$.
Then the second-order correlation function ${g^{(2)}}(0)$ can be expressed as
\begin{equation}
	{g^{(2)}}(0) \approx \frac{{2|{C_{200}}{|^2}}}{{|{C_{100}}{|^4}}}.\label{ana}
\end{equation}

On the other hand, the dynamical evolution of the system can be explored by numerically solving the Lindblad master equation \cite{Scully1997}:
\begin{equation}
	\dot{\hat{\rho}}  =  - i[\hat H,\hat \rho ] - \kappa_c {\cal D}[\hat a]/2 - \kappa_c {\cal D}[\hat b]/2 - {\kappa _m}{\cal D}[\hat m]/2,\label{Lindblad}
\end{equation}
in which $\rho$ denotes the density matrix of the whole system and ${\cal D}[\hat x]\hat \rho  = {{\hat x}^\dag }\hat x\hat \rho  - 2\hat x\hat \rho {{\hat x}^\dag } + \hat \rho {{\hat x}^\dag }\hat x$. When the system approaches its steady state ${\rho }_s$, the numerical results of the second-order correlation function \cite{PhysRev.130.2529} can be expressed as 
\begin{equation}
	{g^{(2)}}(0) = \frac{{\left\langle {{{\hat a}^\dag }{{\hat a}^\dag }\hat a\hat a} \right\rangle }}{{{{\left\langle {{{\hat a}^\dag }\hat a} \right\rangle }^2}}} = \frac{{{\rm{Tr(}}{{\hat a}^\dag }{{\hat a}^\dag }\hat a\hat a{{\hat \rho }_s}{\rm{)}}}}{{{{{\rm{[Tr(}}{{\hat a}^\dag }\hat a{{\hat \rho }_s}{\rm{)]}}}^2}}}.\label{num}
\end{equation}
Typically, ${g^{(2)}}(0) < 1$ characterizes photon antibunching with sub-Poissonian statistics, whereas ${g^{(2)}}(0) > 1$ corresponds to photon bunching with super-Poissonian statistics. In particular, the limit ${g^{(2)}}(0) \to 0$ signifies the occurrence of perfect photon blockade, where only a single photon can be excited in the CCW cavity mode $a$.

Next, we discuss the  optimal conditions for photon blockade in mode $a$. When ${C_{200}}$, the two-photon states are suppressed owing to quantum destructive interference between different transition pathways, resulting in a strong antibunching phenomenon, i.e. photon blockade. Therefore, by solving the numerator of Eq.~(\ref{ana}) to zero, the optimal condition can be obtained as
\begin{equation}
	E{e^{i\phi }} = \frac{{ {V^2}{Z^2}Q}}{{P({{\tilde \Delta }_c}g_a^4 - 2J{g_a}g_a^3 + {K_2}g_a^2 + {K_1}{g_a} + {K_0})}},\label{Ephi}
\end{equation}
which is equivalent to the following two equations:
\begin{equation}
\begin{aligned}
	{E_{{\rm{opt}}}} &= {\rm{Abs}}[\frac{{ {V^2}{Z^2}Q}}{{P({{\tilde \Delta }_c}g_a^4 - 2J{g_b}g_a^3 + {K_2}g_a^2 + {K_1}{g_a} + {K_0})}}],\\
	{\phi _{{\rm{opt}}}} &= {\rm{arg}}[\frac{{ {V^2}{Z^2}Q}}{{P({{\tilde \Delta }_c}g_a^4 - 2J{g_b}g_a^3 + {K_2}g_a^2 + {K_1}{g_a} + {K_0})}}].\label{opt}
\end{aligned}
\end{equation}

\begin{figure*}
	\includegraphics[width=17.2cm, height=5.48cm,, clip]{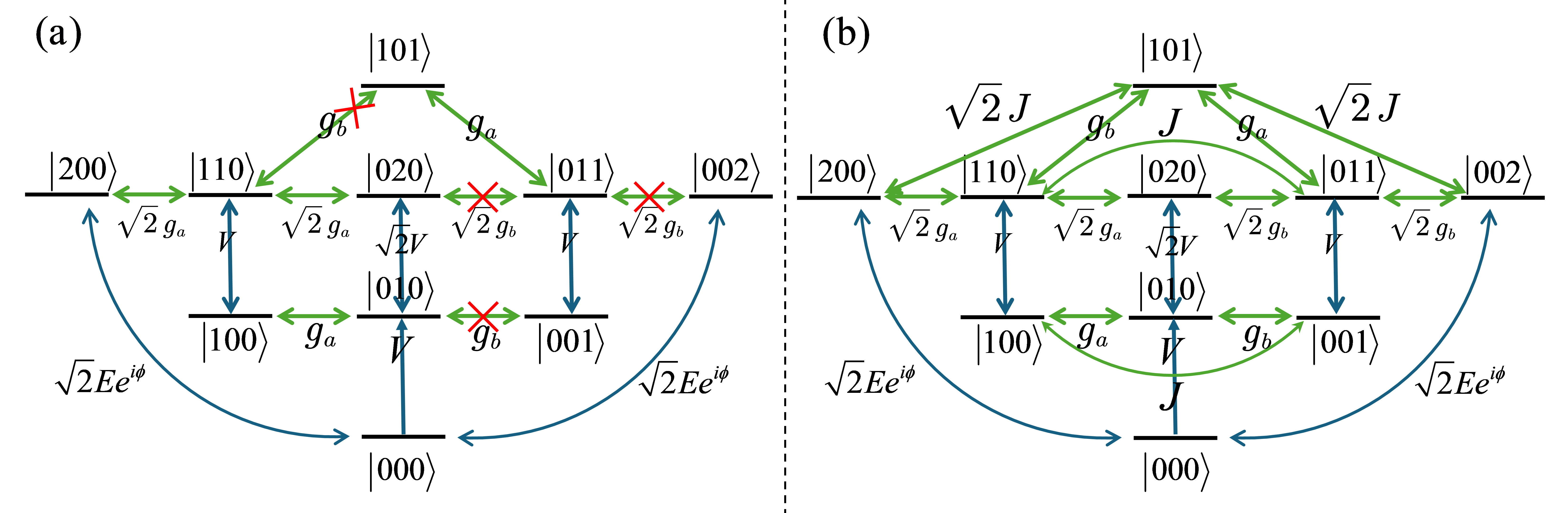}
	\caption{(Color online) Schematic diagram of the energy-level transition paths for (a) $J=0$ and (b) $J\neq0$.}\label{Fig4}
\end{figure*}
\begin{figure}
		\includegraphics[width=8.6cm, height=6.675cm, clip]{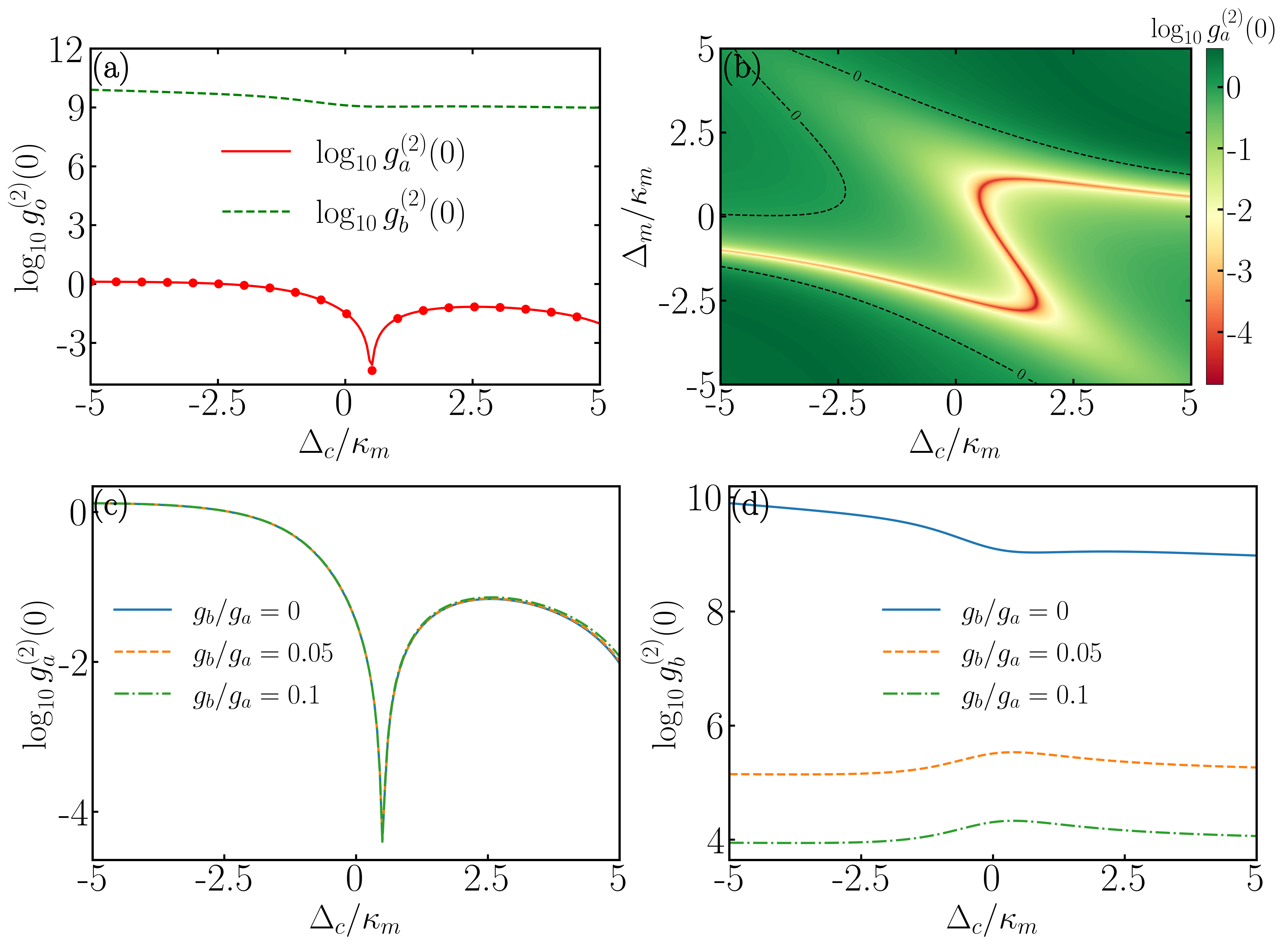}
	\caption{(Color online) Logarithmic plots of the second-order correlation functions ${\log _{10}}g_a^{(2)}(0)$ and ${\log _{10}}g_b^{(2)}(0)$ as functions of the detuning $\Delta_c/\kappa_m$ with $E=E_{{\rm{opt}}}$. (b) ${\log _{10}}g_a^{(2)}(0)$ as a function of  the detunings $\Delta_c/\kappa_m$ and $\Delta_m/\kappa_m$. (c) ${\log _{10}}g_a^{(2)}(0)$ and (d) ${\log _{10}}g_b^{(2)}(0)$ versus  $\Delta_c/\kappa_m$ for different coupling ratios $g_b/g_a$ = 0, 0.05, and 0.1. All other parameters are the same as in Fig.~\ref{Fig2}(c).}\label{Fig5}
\end{figure}

\section{Results and Discussion}\label{Sec3}
In this section, photon blockade in the two modes is analyzed by plotting the logarithm of the second-order correlation function. The logarithmic second-order correlation functions of the two modes are presented in Fig.~\ref{Fig2}(a). As shown, mode $b$ maintains photon bunching over the detuning range from $-5\kappa_m$ to $5\kappa_m$. In contrast, mode $a$ exhibits pronounced photon antibunching at $\Delta_m = 0.5\kappa_m$. These results indicate that, under two-photon driving from both the left and right ports, the counterclockwise mode is effectively suppressed at a specific detuning, leading to a single-photon emission regime, whereas the clockwise mode supports two-photon emission. It can be seen that no green triangular markers appear in Fig.~\ref{Fig2}(a). This is because the probability amplitude $C_{001}$ in Eq.~\ref{solofproapmp} vanishes when $J = g_a = 0$, resulting in the absence of an analytical curve for the second-order correlation function. When $J \neq 0$, the analytical solution of $\log_{10} g_b^{(2)}(0)$ can be observed in Fig.~\ref{Fig8}, showing good agreement between the analytical and numerical results. Figure~\ref{Fig2}(b) shows the mean photon numbers of modes $a$ and $b$. The mean photon number of the clockwise mode $b$ remains nearly unchanged, whereas that of the counterclockwise mode $a$ exhibits a pronounced maximum at $\Delta_m = 0.5\kappa_m$. As $g_a$ increases, the logarithmic second-order correlation functions $\log_{10} g_a^{(2)}(0)$ and $\log_{10} g_b^{(2)}(0)$ as functions of $\Delta_m$ are presented in Figs.~(a) and (d), respectively. When $g_b$ is set to $0.1 g_a$, the second-order correlation function of mode $b$ decreases but remains significantly greater than 1, while that of mode $a$ shows no noticeable change. This demonstrates that the proposed scheme is robust against imperfect chirality.

\begin{figure}
	\includegraphics[width=8.6cm, height=6.675cm, clip]{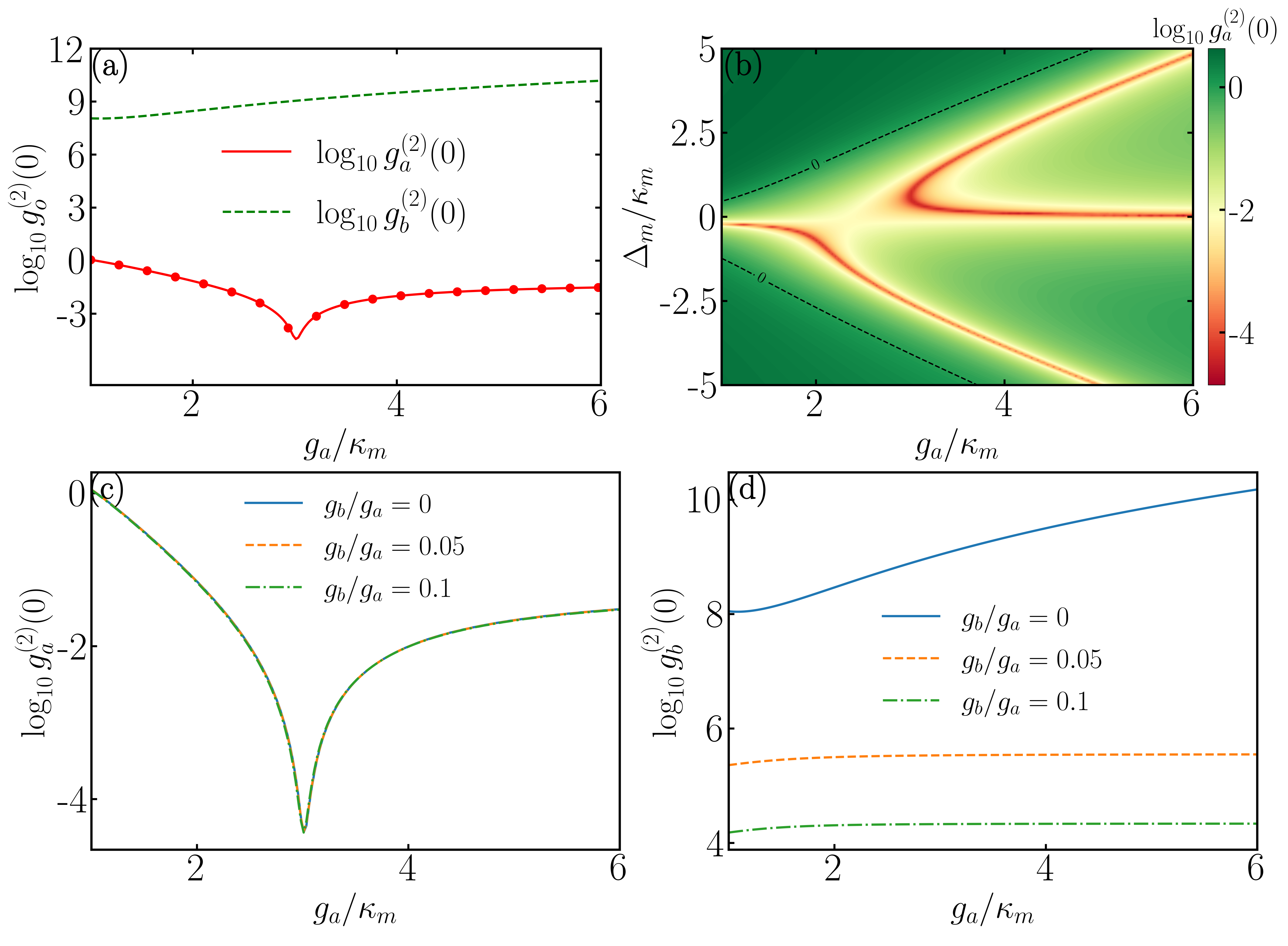}
	\caption{(Color online) Dependence of the logarithmic second-order correlation functions ${\log _{10}}g_a^{(2)}(0)$ and ${\log _{10}}g_b^{(2)}(0)$ on the coupling strength $g_a/\kappa_m$ with $E=E_{{\rm{opt}}}$. (b) ${\log _{10}}g_a^{(2)}(0)$ versus the coupling strength $g_a/\kappa_m$ and the detuning $\Delta_m/\kappa_m$. (c) ${\log _{10}}g_a^{(2)}(0)$ and (d) ${\log _{10}}g_b^{(2)}(0)$ as functions of $g_a/\kappa_m$ for different coupling ratios $g_b/g_a$ = 0, 0.05, and 0.1. The other parameters are the same as in Fig.~\ref{Fig2}(c).}\label{Fig6}
\end{figure}
To verify the optimal conditions for nonreciprocal photon blockade, we plot the logarithmic second-order correlation function $\log_{10} g_a^{(2)}(0)$ of mode $a$ in Figs.~\ref{Fig3}(a) and~\ref{Fig3}(b) as functions of the phase $\phi$, and of the driving strength $E$ and detuning $\Delta_m$, respectively. The orange dashed lines indicate the optimal parameter values for realizing nonreciprocal photon blockade. As shown in the figures, when $E = E_{\mathrm{opt}}$ or $\phi = \phi_{\mathrm{opt}}$, one can always find a corresponding value of $\phi$ or $E$ for which the second-order correlation function is significantly smaller than 1 . As can be seen from Fig.~\ref{Fig3}(a), when $E = E_{\mathrm{opt}}$, the optimal value of the phase $\phi$ is not unique. Instead, there exist multiple values of $\phi$ (including $\phi = -0.41\pi$) at which the second-order correlation function of mode $a$ reaches a minimum. This feature explains why two minima appear in Fig.~\ref{Fig2}(a) when $\phi$ is fixed at $-0.41\pi$. Furthermore, under the condition $\Delta_c = \Delta_m = 0.5\kappa_m$, we plot $\log_{10} g_a^{(2)}(0)$ as functions of the phase $\phi$ and the driving strength $E$ in Fig.~\ref{Fig3}(c). It can be observed that when $E = 0.0002\kappa_m$ and $\phi = -0.41\pi$, $\log_{10} g_a^{(2)}(0)$ reaches its minimum, which is consistent with the optimal condition. In the following analysis, we adopt these optimal parameters $E$ and $\phi$ to further investigate the photon blockade effect.

Below we describe the physical mechanism underlying directional photon blockade. As shown in Fig.~\ref{Fig4}(a), when the coupling between the two cavity modes is neglected, the chiral cavity–magnon interaction ensures that $g_a \neq 0$ while $g_b = 0$. Consequently, all transition paths induced by the cavity–magnon coupling $g_b$ are suppressed, as indicated by the red crosses in the figure. As a result, only a single remaining transition path leading to the state $|002\rangle$ survives. Under otherwise identical parameters, this implies that the $a$ mode can exhibit UPB through destructive interference between different excitation pathways, whereas the $b$ mode cannot due to the absence of such interfering paths. For completeness, the full set of transition paths, including the coupling between the two cavity modes, is presented in Fig.~\ref{Fig4}(b).

Figure~\ref{Fig5} illustrates the evolution of the logarithmic second-order correlation functions, $\log_{10} g_a^{(2)}(0)$ and $\log_{10} g_b^{(2)}(0)$, as functions of the detuning $\Delta_c$. At $\Delta_c = 0.5\kappa_m$, which corresponds to the parameters identified in Fig.~\ref{Fig3}, mode $a$ exhibits a pronounced minimum, indicating clear photon antibunching. In contrast, the correlation function of mode $b$ remains significantly greater than 1, thereby demonstrating nonreciprocal photon blockade. To further elucidate the dependence of the correlation function on the two detunings, we present a contour plot of $\log_{10} g_a^{(2)}(0)$ in the $\Delta_c$–$\Delta_m$ parameter space. A distinct red trajectory is observed, highlighting the region where the correlation function of mode $a$ is much smaller than 1. Since the optimal parameters are located at $\Delta_c = \Delta_m = 0.5\kappa_m$, the photon blockade effect is clearly identified at this point. Furthermore, Figs.~\ref{Fig5}(c) and~\ref{Fig5}(d) show $\log_{10} g_a^{(2)}(0)$ and $\log_{10} g_b^{(2)}(0)$ as functions of $\Delta_c$ for different ratios of $g_b/g_a$. The results indicate that, at $\Delta_c = 0.5\kappa_m$, a finite $g_b/g_a$ ratio arising from imperfect chiral coupling has a negligible effect on the antibunching behavior of mode $a$. That is, mode $a$ continues to exhibit pronounced photon blockade at the optimized detuning. By contrast, although $\log_{10} g_b^{(2)}(0)$ decreases for the three considered values of $g_b/g_a$, it remains well above unity. Consequently, nonreciprocal photon blockade is clearly established at $\Delta_c = 0.5\kappa_m$.

Figure~\ref{Fig6} investigates the influence of the coupling strength $g_a$ on the photon blockade behavior of the system. In Fig.~\ref{Fig6}(a), the logarithmic second-order correlation function $\log_{10} g_a^{(2)}(0)$ is plotted as a function of $g_a$. Similar to Fig.~\ref{Fig5}(a), the correlation function reaches a minimum at $g_a = 3\kappa_m$, corresponding to the optimal value identified in Fig.~\ref{Fig3}(c), where nonreciprocal photon blockade is observed. Figure~\ref{Fig6}(b) shows the dependence of $\log_{10} g_a^{(2)}(0)$ on both $g_a$ and the detuning $\Delta_m$. The results indicate that, instead of a single trajectory of minima, two prominent low-value branches emerge. In particular, at $g_a = 3\kappa_m$, there exist two corresponding detuning values at which the correlation function reaches its minimum. Notably, for $g_a > 3\kappa_m$, mode $a$ exhibits three distinct detuning points where pronounced photon blockade occurs. Furthermore, an analysis of different $g_b/g_a$ ratios shows that the antibunching behavior of mode $a$ is robust against imperfect chiral coupling, with the corresponding curves nearly overlapping. In contrast, the correlation function of mode $b$ increases with $g_a$ and exhibits clear separation for different $g_b/g_a$ ratios. Although $\log_{10} g_b^{(2)}(0)$ decreases as $g_b/g_a$ increases, it remains well above unity. These results demonstrate that stable nonreciprocal photon blockade is maintained over the considered parameter range.

The origin of the multiple minimum branches in Fig.~\ref{Fig6}(b) can be understood from the analytical optimal condition given by Eq.~\ref{opt}. In the numerical calculations, the driving strength is fixed at its optimal value $E = E_{\mathrm{opt}}$, and the phase is chosen as $\phi = -0.41\pi$, as identified from Fig.~\ref{Fig3}(a). According to Eq.~\ref{opt}, the optimal photon blockade condition is determined by the phase relation $\mathfrak{E} _1 e^{i\theta_1}/{\mathfrak{E} _2 e^{i\theta_2}}= -0.41\pi$. This relation can be equivalently expressed as a constraint on the phase difference, namely $\phi_1 - \phi_2 = -0.41\pi$. Under this condition, substituting $g_a$ and $\Delta_m$ into Eq.~\ref{opt} yields the parameter combinations that satisfy the optimal photon blockade condition, which correspond precisely to the red low-value trajectories in Fig.~\ref{Fig6}(b). For relatively large $g_a$, the resulting equation becomes a higher-order polynomial in $\Delta_m$, where the numerator contains terms up to second order while the denominator includes terms up to fourth order. Consequently, multiple solutions for $\Delta_m$ may arise. When combined with the above phase constraint, the number of allowed solutions is further restricted. As a result, for a fixed coupling strength $g_a$, up to three distinct detuning values $\Delta_m$ can satisfy the optimal condition, giving rise to the multiple minima observed in Fig.~\ref{Fig6}(b). This analytical insight also accounts for the multiple solution behavior observed in Fig.~\ref{Fig5}(b).

\begin{figure}
	\includegraphics[width=8.6cm, height=3.337cm, clip]{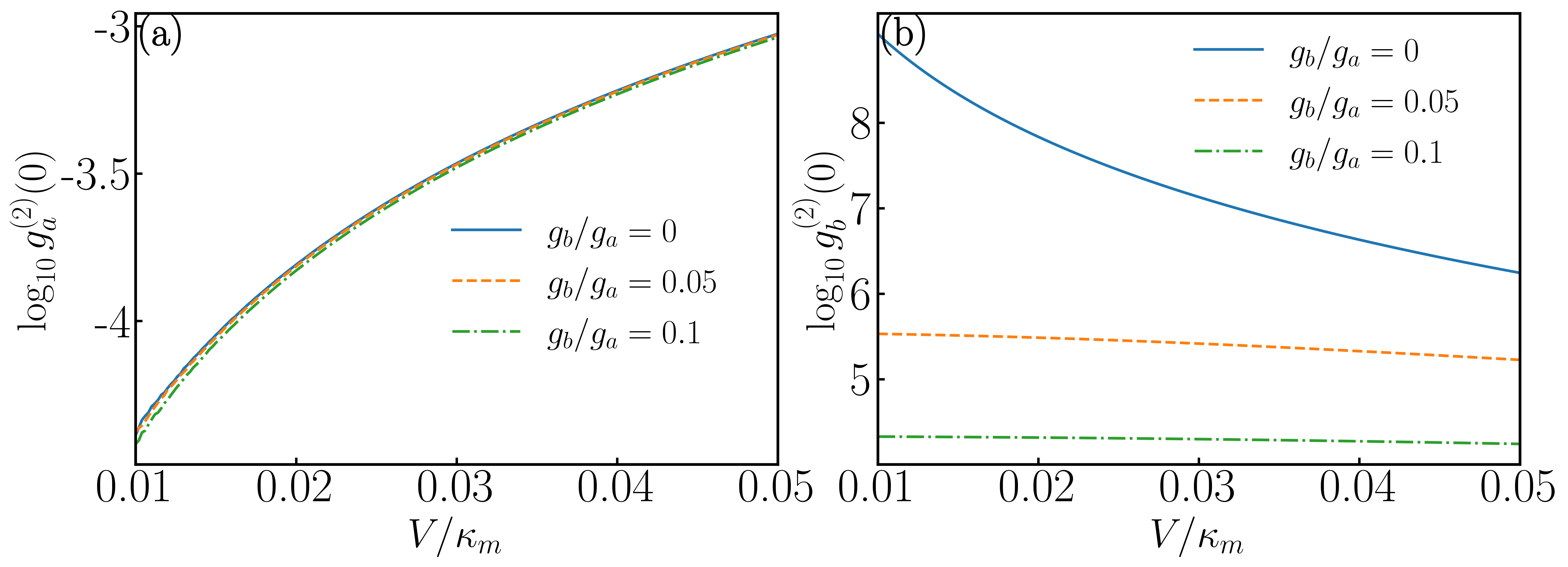}
	\caption{(Color online) (a) Logarithmic second-order correlation function ${\log _{10}}g_a^{(2)}(0)$ and (b) ${\log _{10}}g_b^{(2)}(0)$ versus the driving strength $V/\kappa_m$ for the different coupling ratios $g_b/g_a$ = 0, 0.05, and 0.1.}\label{Fig7}
\end{figure}

\begin{figure}
	\includegraphics[width=8.6cm, height=3.337cm, clip]{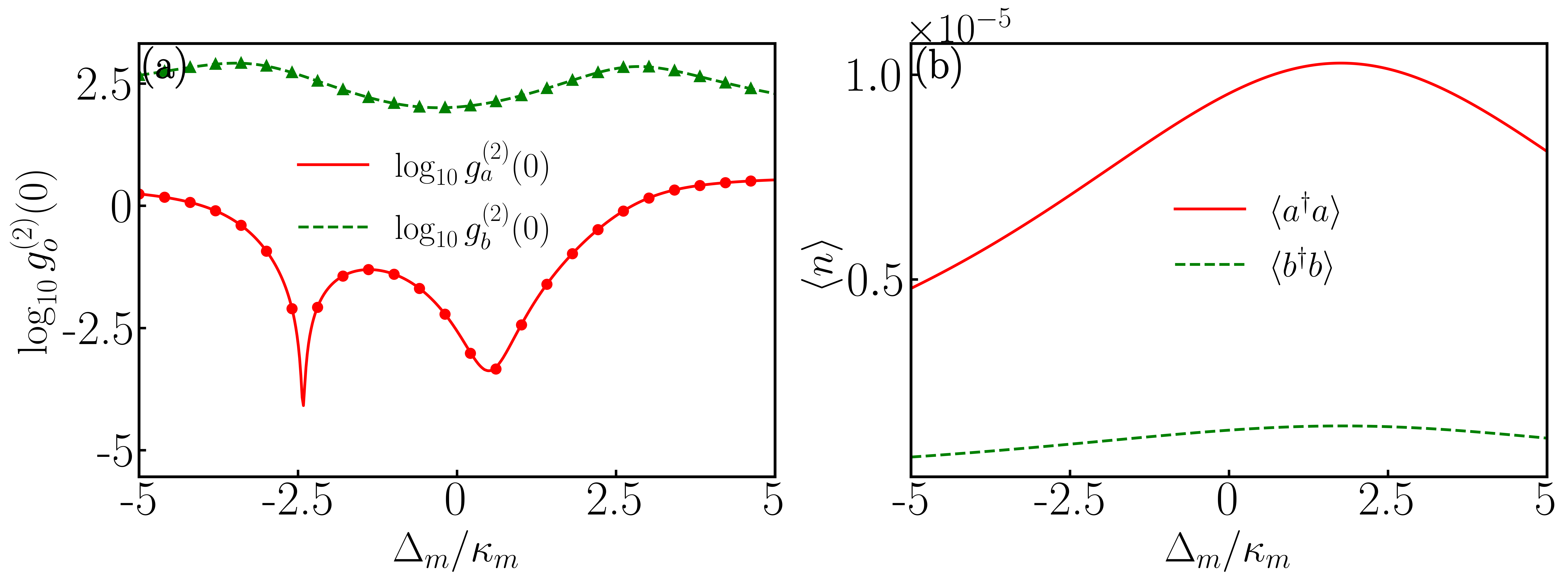}
	\caption{(Color online) (a) Logarithmic  second-order correlation functions ${\log _{10}}g_a^{(2)}(0)$ and ${\log _{10}}g_b^{(2)}(0)$,  and (b) the mean photon number of mode $a$ and mode $b$, are plotted as functions of the detuning $\Delta_m/\kappa_m$ with $J= 0.5\kappa_m$.}\label{Fig8}
\end{figure}

To investigate the influence of the driving strength on the second-order correlation functions, Fig.~\ref{Fig7} illustrates the evolution of ${\log_{10}} g_a^{(2)}(0)$ and ${\log_{10}} g_b^{(2)}(0)$ as functions of the driving strength $V$ for different ratios of $g_b/g_a$. As $V$ increases from $0.01\kappa_m$ to $0.05\kappa_m$, the value of ${\log_{10}} g_a^{(2)}(0)$ increases by approximately one unit on the logarithmic scale, while ${\log_{10}} g_b^{(2)}(0)$ exhibits varying degrees of reduction. These results indicate that the nonreciprocal photon blockade is maintained within a finite range of driving strengths, even when the driving strength deviates from the optimal value $V = 0.01\kappa_m$ corresponding to the optimized phase $\phi$. Moreover, the curves for mode $a$ nearly overlap for the three different $g_b/g_a$ ratios, indicating that imperfect chiral coupling has a negligible influence on the photon blockade behavior of mode $a$. Consequently, the system exhibits a high degree of robustness against imperfect chirality.

Due to dielectric backscattering from the sphere or the coupling ports~\cite{PhysRevB.89.224407}, the coupling between the two cavity modes is not always ideal (i.e., $J = 0$) and typically acquires a finite value. Accordingly, we set $J = 0.5\kappa_m$ and plot the second-order correlation functions (on a logarithmic scale) as well as the mean photon numbers as functions of the detuning $\Delta_m$ in Fig.~\ref{Fig8}, while keeping all other parameters identical to those used in Fig.~\ref{Fig2}(a). Compared with Fig.~\ref{Fig2}(a), the logarithmic value of the second-order correlation function for mode $b$ is significantly reduced, decreasing from approximately $9$ to about $2.5$. This behavior can be understood from the energy-level diagram shown in Fig.~\ref{Fig4}(b): when $J \neq 0$, the state $|001\rangle$ can be populated through the intermode coupling pathway even in the absence of direct coupling ($g_b = 0$). As a result, mode $b$ is no longer confined to the $|002\rangle$ state, and correspondingly, its mean photon number becomes nonzero, as shown in Fig.~\ref{Fig8}(b). In contrast, the presence of a finite $J$ has a negligible influence on mode $a$. Specifically, at $\Delta_m = 0.5\kappa_m$, although ${\log_{10}} g_a^{(2)}(0)$ increases from below $10^{-4}$ to approximately $10^{-3}$, the system still exhibits strong photon antibunching. Moreover, the mean photon number of mode $a$ shows no noticeable variation. Consequently, robust nonreciprocal photon blockade is preserved at $\Delta_m = 0.5\kappa_m$.

\begin{figure}
	\includegraphics[width=8.6cm, height=3.337cm, clip]{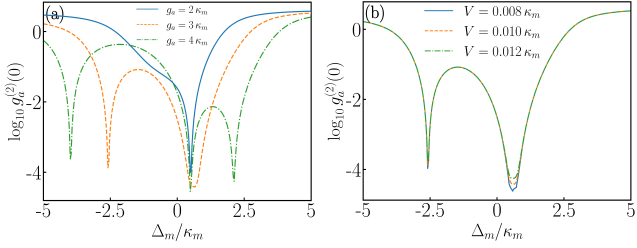}
	\caption{(Color online) Logarithmic plots of the second-order correlation function ${\log _{10}}g_a^{(2)}(0)$ versus the detuning $\Delta_m/\kappa_m$ for (a) the different coupling strengths $g_a/\kappa_m$ = 2, 3, and 4, and (b) the differnet driving strengths $V/\kappa_m$ = 0.008, 0.01, and 0.012. For each set of parameters, the phase $\phi$ is taken as the optimal value at $\Delta_m = 0.5\kappa_m$, and all other parameters are the same as in Fig.~\ref{Fig2}(c).}\label{Fig9}
\end{figure}

To further elucidate this behavior, we plot the logarithmic second-order correlation function of mode $a$ in Fig.~\ref{Fig9} for various coupling strengths $g_a$ and driving strengths $V$. Here, the phase $\phi$ is fixed at its optimal value corresponding to $\Delta_m = 0.5\kappa_m$. As depicted in Fig.~\ref{Fig9}(a), ${\log _{10}}g_a^{(2)}(0)$ exhibits two local minima at $g_a/\kappa_m=3$, whereas one and three distinct dips appear for $g_a/\kappa_m = 2$ and $4$, respectively, in good agreement with the results in Fig.~\ref{Fig6}(b). Notably, at $\Delta_m = 0.5\kappa_m$, ${\log _{10}}g_a^{(2)}(0)$ remains well below 0 even as the coupling strength increases, confirming strong photon antibunching and the persistence of photon blockade in mode $a$. In Fig.~\ref{Fig9}(b), although the optimal phase is derived using $V = 0.01\kappa_m$ via Eq.~\ref{opt}, the minimum of ${\log _{10}}g_a^{(2)}(0)$ at $\Delta_m = 0.5\kappa_m$ shows negligible fluctuation as $V$ varies. This demonstrates that the photon blockade effect is robust against variations in both coupling and driving strengths.

\section{Conclusion}\label{Sec4}
In summary, we propose a scheme for realizing directional photon blockade in a chiral cavity–magnon system. When a YIG sphere is placed inside a torus-shaped microwave cavity, at specific positions, both the CW and CCW modes become circularly polarized, such that only one of them couples to the Kittel mode. Under two-photon driving, this chiral interaction induces unconventional photon blockade exclusively in one propagation direction, thereby enabling directional single-photon emission. Furthermore, the emission direction can be switched on demand by inverting the bias magnetic field. We also investigated the system's performance in the presence of imperfect chiral coupling and intermode backscattering. Our results confirm that directional photon blockade persists under these realistic conditions, highlighting the robustness of the scheme. Given that our proposal relies on established experimental configurations, it holds promise for near-term implementation and offers new perspectives for the development of nonreciprocal quantum devices.

\begin{acknowledgments}
	The authors gratefully acknowledge the project funded by the National Natural Science Foundation of China (Grant No.
	11874251) and Xi’an Science and Technology Plan Project (Grant No. 23KGDW0026-2022).
\end{acknowledgments}

\appendix

\section{REMAINING EXPRESSIONS FOR THE PROBABILITY AMPLITUDES}\label{AppendixA}
In this appendix, we provide the remaining probability-amplitude expressions that were not included in the main text.
\begin{widetext}
\begin{equation}
\begin{aligned}
	{C_{010}} &= \frac{{VM}}{P},\\
	{C_{001}} &= \frac{{VY}}{P},\\
	{C_{110}} &= \frac{{{V^2}ZMQ + {E{e^{i\phi }}P( - {{\tilde \Delta }_c}{{\tilde \Delta }_m}g_a^3 + {A_2}g_a^2 + {A_1}{g_a} + {A_0})}}}{{{P^2}}Q},\\
	{C_{101}} &= \frac{{{V^2}ZYQ + {E{e^{i\phi }}P({{\tilde \Delta }_m}(g_a^3{g_b} + {g_a}g_b^3) + 4Jg_a^2g_b^2 + {F_2}(g_a^2 + g_b^2) + {F_1}{g_a}{g_b} + {F_0})}}}{{{P^2}}Q},\\
	{C_{011}} &= \frac{{{V^2}MYQ + {E{e^{i\phi }}P( - {{\tilde \Delta }_c}{{\tilde \Delta }_m}g_b^3 + {B_2}g_b^2 + {B_1}{g_b} + {B_0})}}}{{{P^2}}Q},\\
	{C_{020}} &= \frac{{{V^2}{M^2}Q + {E{e^{i\phi }}P({{\tilde \Delta }_c}(g_a^4 + g_b^4) - 2J(g_a^3{g_b} + {g_a}g_b^3) + 2{{\tilde \Delta }_c}g_a^2g_b^2 + {R_1}(g_a^2 + g_b^2) + {R_0}{g_a}{g_b})}}}{{\sqrt 2{P^2}}Q},\\
	{C_{002}} &= \frac{{{V^2}{Y^2}Q - {E{e^{i\phi }}P({{\tilde \Delta }_c}g_b^4 - 2J{g_a}g_b^3 + {L_2}g_b^2 + {L_1}{g_b} + {L_0})}}}{{\sqrt 2{P^2}}Q},\label{Asolofproapmp}
\end{aligned}
\end{equation}
\end{widetext}
where  $M =  - {J^2} + \tilde \Delta _c^2$; $Y = {g_a}J - {g_b}{{\tilde \Delta }_c}$; ${A_2} = {g_b}J( - 2{{\tilde \Delta }_c} + {{\tilde \Delta }_m})$, ${A_1} = {{\tilde \Delta }_c}{{\tilde \Delta }_m}( - g_b^2 + 2({J^2} + {{\tilde \Delta }_c}({{\tilde \Delta }_c} + {{\tilde \Delta }_m})))$, ${A_0} = {g_b}J(2{{\tilde \Delta }_c} + {{\tilde \Delta }_m})(g_b^2 - 2{{\tilde \Delta }_c}{{\tilde \Delta }_m})$; ${F_2} =  - J(2J\tilde \Delta _c^2 + 2{{\tilde \Delta }_c}{{\tilde \Delta }_m} + \tilde \Delta _m^2)$, ${F_1} = {J^2}(4{{\tilde \Delta }_c} - 2{{\tilde \Delta }_m}) - 2{{\tilde \Delta }_c}{{\tilde \Delta }_m}({{\tilde \Delta }_c} + {{\tilde \Delta }_m})$, ${F_0} = 2J{{\tilde \Delta }_c}{{\tilde \Delta }_m}( - {J^2} + {({{\tilde \Delta }_c} + {{\tilde \Delta }_m})^2})$; ${B_2} = {g_a}J( - 2{{\tilde \Delta }_c} + {{\tilde \Delta }_m})$, ${B_1} = {{\tilde \Delta }_c}{{\tilde \Delta }_m}( - g_a^2 + 2({J^2} + {{\tilde \Delta }_c}({{\tilde \Delta }_c} + {{\tilde \Delta }_m})))$, ${B_0} = {g_a}J(2{{\tilde \Delta }_c} + {{\tilde \Delta }_m})(g_a^2 - 2{{\tilde \Delta }_c}{{\tilde \Delta }_m})$; ${R_1} =  - 2{J^2}{{\tilde \Delta }_c} - 2\tilde \Delta _c^3 - 2\tilde \Delta _c^2{{\tilde \Delta }_m}$, ${R_0} = 8J\tilde \Delta _c^2 + 4J{{\tilde \Delta }_c}{{\tilde \Delta }_m}$; ${L_2} = {J^2}{{\tilde \Delta }_m} + (g_a^2 - {{\tilde \Delta }_c}({{\tilde \Delta }_c} + {{\tilde \Delta }_m}))(2{{\tilde \Delta }_c} + {{\tilde \Delta }_m})$, ${L_1} = 2{g_a}J(g_a^2 + 2\tilde \Delta _c^2)$, ${L_0} = g_a^4({{\tilde \Delta }_c} + {{\tilde \Delta }_m}) - g_a^2({J^2}{{\tilde \Delta }_m} + {{\tilde \Delta }_c}({{\tilde \Delta }_c} + {{\tilde \Delta }_m})(2{{\tilde \Delta }_c} + 3{{\tilde \Delta }_m})) + 2\tilde \Delta _c^2{{\tilde \Delta }_m}( - J + {{\tilde \Delta }_c} + {{\tilde \Delta }_m})(J + {{\tilde \Delta }_c} + {{\tilde \Delta }_m})$.


\bibliography{Reference}
\end{document}